\documentclass[a4paper,12pt]{article}
\usepackage[pctex32]{graphics}
\usepackage{amssymb,amsmath}
\usepackage{amsmath,amssymb}
\usepackage{latexsym}
\usepackage{epsfig}
\usepackage[english]{babel}

\newcommand{\be}{\begin{equation}}
\newcommand{\ee}{\end{equation}}
\newcommand{\ba}{\begin{eqnarray}}
\newcommand{\ea}{\end{eqnarray}}

\begin{document}

\begin{titlepage}

\vspace{5mm}

\begin{center}

{\Large \bf Comment on Ricci dark energy in Chern-Simons modified
gravity}

\vskip .6cm

\centerline{\large
 Yun Soo Myung$^{a}$}

\vskip .6cm

{Institute of Basic Science and Department  of Computer Simulation,
\\Inje University, Gimhae 621-749, Korea \\}

\end{center}

\begin{center}
\underline{Abstract}
\end{center}

We revisit  Ricci dark energy in Chern-Simons modified gravity.  As
far as the cosmological evolution, this is nothing but the Ricci
dark energy with a minimally coupled  scalar without potential which
means that the role of Chern-Simons term is suppressed. Using the
equation of state parameter, this model is similar to the modified
Chaplygin gas
 model only when two are around the de Sitter universe deriving by the cosmological constant in the future.
 However, two past evolutions are different.
  \vskip .6cm


\vskip 0.8cm

\vspace{15pt} \baselineskip=18pt

\noindent $^a$ysmyung@inje.ac.kr \\

\thispagestyle{empty}
\end{titlepage}

\newpage
%%%%%%%%%%%%%%%%%%%%%%%%%%%%%%%%%%%%%%%%%%%%%%%%%%%%%%%%%%%%%%%%%%

Recently, the authors~\cite{Silva:2013yaa} have investigated the
Ricci dark energy model in the dynamic Chern-Simons modified gravity
which states  that its cosmological evolution  is similar to  that
displayed by the modified Chaplygin gas model.

In this Comment, we wish to draw the reader two important issues:
One is that as far as the cosmological evolution, this model  is
nothing but the Ricci dark energy with a minimally coupled  scalar
without potential where the role of Chern-Simons term is suppressed
totally. The other is that using the equation of state parameter,
this model is similar to the modified Chaplygin gas
 model only when two make  turnaround of de Sitter universe deriving by the cosmological constant in the future.
In general, two provide different evolutions.

We start with  the dynamic Chern-Simons modified gravity action with
Ricci dark energy~\cite{Silva:2013yaa}
 \be\label{gmg}
 S=\frac{1}{16\pi G}\int d^4x~\sqrt{-g}\left[R-\frac{\tilde{\theta}}{4}~{}^*RR-\frac{1}{2}\partial^\mu \tilde{\theta} \partial_\mu \tilde{\theta}+V(\tilde{\theta})\right]+S_{\rm RDE},
 \ee
where ${}^*RR$ is the Pontryagin term,  $\tilde{\theta}$ is a
dynamical scalar and $S_{\rm RDE}$  is the action to give the Ricci
dark energy. Here, for simplicity, one chooses
$V(\tilde{\theta})=0$. Their equations are given by
 \ba \label{ein1}
 G_{\mu\nu}+C_{\mu\nu}&=&8\pi G T_{\mu\nu},\\
\label{ein2} \nabla^2\tilde{\theta}&=&-\frac{1}{64\pi}{}^*RR,
 \ea
where $G_{\mu\nu}$ is the Einstein tensor, $C_{\mu\nu}$ is the
Cotton tensor from the Chern-Simons term
``$\tilde{\theta}~{}^*RR$''-term. The energy-momentum tensor is
given by \be T_{\mu\nu}=T^{\rm RDE}_{\mu\nu}+T^{\rm
\tilde{\theta}}_{\mu\nu} \ee with  \be T_{\mu\nu}^{\rm
RDE}=(\rho_{\rm RDE}+p_{\rm RDE})u_\mu u_\nu+p_{\rm RDE} g_{\mu\nu}
\ee and \be T_{\mu\nu}^{\rm \tilde{\theta}}=\partial_\mu\tilde{
\theta}
\partial_\nu \tilde{\theta}-\frac{1}{2}g_{\mu\nu}\partial_\mu \tilde{\theta}
\partial_\mu \tilde{\theta}. \ee
Applying $\nabla^\mu$ to (\ref{ein1}) leads to the conservation-law
for $T^{\mu\nu}$ as \be \label{con-l}\nabla_\mu T^{\mu\nu}=0, \ee
which plays an important role in the cosmological evolution.

 In this work, we consider the flat
Friedmann-Robertson-Walker (FRW) spacetimes \be \label{sle}
ds^2_{\rm FRW}=g_{\mu\nu}dx^\mu dx^\nu=-dt^2+a(t)^2\Big(dr^2+r^2
d\theta^2+r^2\sin^2\theta d\phi^2\Big). \ee From (00)-component of
(\ref{ein1}), we have the Friedmann equation with $G=1$ \be
\label{fried}
H^2=\alpha(2H^2+\dot{H})+\frac{4\pi}{3}\dot{\tilde{\theta}}. \ee In
deriving (\ref{fried}), we used \be
G_{00}=3H^2,~C_{00}=0,~~\rho_{\rm
RDE}=-\frac{\alpha}{16\pi}R=\frac{6\alpha}{16\pi}(2H^2+\dot{H}),~~T^\theta_{00}=\frac{1}{2}\dot{\tilde{\theta}}.
\ee We note here that the Cotton tensor $C_{\mu\nu}$ vanishes for
the FRW metric (\ref{sle}), implying that
``$\tilde{\theta}~{}^*RR$''-term does not derive any cosmological
evolution. Therefore, the whole equations reduce to those of  the
Ricci dark energy model with a minimally coupled scalar
$\tilde{\theta}$.

Because of ${}^*RR=0$ for the FRW metric (\ref{sle}), equation
(\ref{ein2}) leads to the  conservation-law for $\tilde{\theta}$ \be
\label{con-t}\nabla^2\tilde{\theta}=0 \to
\ddot{\tilde{\theta}}+3H\dot{\tilde{\theta}}=0, \ee whose solution
is given by \be \label{t-the}\dot{\tilde{\theta}}=\frac{C}{a^3}. \ee
Finally, the conservation-law (\ref{con-l}) provides  \be
\dot{\rho}_{\rm RDE}+3H(\rho_{\rm RDE}+p_{\rm
RDE})+\ddot{\tilde{\theta}}+3H\dot{\tilde{\theta}}=0, \ee while
using the conservation-law for $\tilde{\theta}$ (\ref{con-t}), it
leads to the conservation-law for the Ricci dark energy \be
\label{ricci}\dot{\rho}_{\rm RDE}+3H(\rho_{\rm RDE}+p_{\rm RDE})=0.
\ee Eqs.(\ref{fried}), (\ref{con-t}), and (\ref{ricci}) state that
whole evolution equations amount to the Ricci dark energy with a
minimally coupled scalar. This is because the Chern-Simons term of
$\tilde{\theta}~{}^*RR$ does not contribute to the cosmological
evolution. However, the cosmological perturbation will distinguish
between Ricci dark energy in Chern-Simons modified gravity and Ricci
dark energy with a minimally coupled scalar
$\tilde{\theta}$~\cite{Alexander:2009tp}. At this stage, we wish to
mention that the conservation-law (\ref{ricci}) might be not useful
to see the cosmological evolution because the Friedmann equation
(\ref{fried}) does not belong to the standard one due to $\rho_{\rm
RDE}$.

Plugging (\ref{t-the}) into (\ref{fried}) and then, expressing it in
terms of scale factor $a$ leads to~\cite{Silva:2013yaa} \be \alpha
\frac{\ddot{a}}{a}+(\alpha-1)\Big(\frac{\dot{a}}{a}\Big)^2+\frac{\beta}{a^6}=0\ee
with $\beta=4\pi C^2/3$. For $\alpha \simeq 1/2$, this was solved
for $a(t)$ to be \be \label{scale}
a(t)=\Big(\frac{2\beta}{3c_1}\Big)^{1/6}\sinh^{1/3}\Big[3\sqrt{c_1}
t\Big],\ee where $c_1$ is an undetermined integration constant.

The authors~\cite{Silva:2013yaa} insisted that there is a
correspondence between the Ricci dark energy in Chern-Simons
modified gravity and the modified Chaplygin gas model because the
solution (\ref{scale}) was also found in the modified Chaplygin gas
model.  Aside from the fact that Ricci dark energy in Chern-Simons
modified gravity reduces to Ricci dark energy with a minimally
coupled scalar, the similarity between two is very restrictive and
it is limited to the  de Sitter phase derived by the cosmological
constant in the future. Therefore, discovering (\ref{scale}) is not
sufficient to confirm the correspondence between two models.   In
order to show this explicitly, we rewrite (\ref{fried}) as the
first-order inhomogeneous equation
 for $H^2$  with $x=\ln a$ instead of scale factor $a$~\cite{Kim:2008ej} \be \label{first-i}
\frac{dH^2}{dx}+\Big(4-\frac{2}{\alpha}\Big)H^2=-\frac{2}{3\alpha}\rho_{\tilde{\theta}}
\ee with \be \rho_{\tilde{\theta}}=\rho_{\tilde{\theta}
0}e^{-6x},~~\rho_{\tilde{\theta} 0}=\pi C^2 =\frac{3\beta}{4}. \ee A
new variable $x = \ln[a/a_0]$ with $a_0=1$  ranges from $-\infty$ to
$\infty$ which  includes the present $ x = 0$ at $a = a_0
$.
  It
is important to note  that $\rho_{\tilde{\theta}}$ plays a role of
the stiff matter with its equation of state
$\omega_{\tilde{\theta}}=1$. Eq.(\ref{first-i}) could be integrated
to give the standard Friedmann equation with a positive integration
constant $\tilde{c}_1$ as \be \label{sfeq} H^2=\frac{\rho_t}{3} \ee
with \be \label{rhot}\rho_t=\frac{\rho_{\tilde{\theta}0}
e^{-6x}}{\alpha(\alpha+1)}+3\tilde{c}_1e^{-(4-\frac{2}{\alpha})}.
\ee The total energy density is divided into two parts as \ba
\label{afried} \rho_t&=&\rho_{\tilde{\theta}
0}e^{-6x}+\Big\{\frac{1-\alpha(\alpha+1)}{\alpha(\alpha+1)}\rho_{\tilde{\theta}
0}e^{-6x}+3\tilde{c}_1 e^{-(4-\frac{2}{\alpha})x}\Big\} \\
\label{afried1}&\equiv& \rho_{\tilde{\theta}}+\tilde{\rho}_{\rm
RDE},\ea where a new scaled Ricci dark energy density is given by
\be \label{nrde}\tilde{\rho}_{\rm
RDE}=\frac{1-\alpha(\alpha+1)}{\alpha(\alpha+1)}\rho_{\tilde{\theta}
0}e^{-6x}+3\tilde{c}_1 e^{-(4-\frac{2}{\alpha})x}. \ee For
$\alpha(\alpha+1)<1$, one finds that $\tilde{\rho}_{\rm RDE}>0$.
Without the scalar $\tilde{\theta}$, the pure Ricci dark energy
density is given only by the second term in (\ref{nrde}) when
expressing the standard Friedmann equation like
(\ref{sfeq})~\cite{Kim:2008ej}. In this case, its equation of state
is given by \be \omega_{\rm
RDE}=\frac{1}{3}\Big(1-\frac{2}{\alpha}\Big) \ee which shows that
for $\alpha <1$, it describes the dark energy-dominated universe.
Also, we note that the energy density $\rho_{\tilde{\theta}}$ in
(\ref{afried1}) satisfies the conservation-law as \be
\tilde{p}_{\tilde{\theta}}=-\tilde{\rho}_{\tilde{\theta}}-\frac{1}{3}\frac{d\tilde{\rho}_{\tilde{\theta}}}{dx}\ee
for $\tilde{p}_{\tilde{\theta}}=\omega_{\tilde{\theta}}
\tilde{\rho}_{\tilde{\theta}}$ with $\omega_{\tilde{\theta}}=1$,
which indicates that the pure kinetic term of $\tilde{\theta}$ plays
a role of the stiff matter.

Substituting $\tilde{\rho}_{\rm RDE}$ into the conservation-law, \be
\tilde{p}_{\rm RDE}=-\tilde{\rho}_{\rm
RDE}-\frac{1}{3}\frac{d\tilde{\rho}_{\rm RDE}}{dx}, \ee we obtain
the Ricci dark energy pressure \be \tilde{p}_{\rm RDE}=
\frac{1-\alpha(\alpha+1)}{\alpha(\alpha+1)}\rho_{\theta
0}e^{-6x}+(1-\frac{2}{\alpha})\tilde{c}_1e^{-(4-\frac{2}{\alpha})x}.\ee
%%%%%%%%%%%%%%%%%%%%%%%%%%%%%%%%%%%%%%%%%%%%%%%%%%%%%%%%%%%%%%%%%%%%%%%%%
\begin{figure*}[t!]
   \centering
   \includegraphics{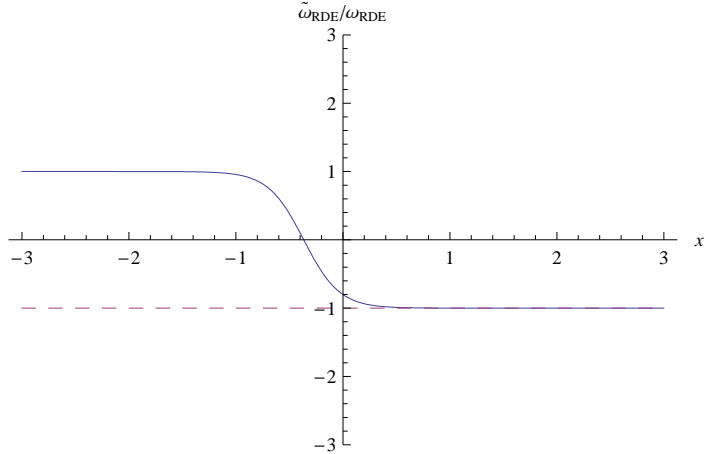}
\caption{Two equation of state parameters as functions of $x$ for
$\rho_{\tilde{\theta}0}=\tilde{c}_1=1$ and $\alpha=1/2$. $x =
\ln[a/a_0]$ with $a_0=1$  ranges from $-\infty$ to $\infty$ which
includes the present $ x = 0$ at $a = a_0$. $\omega_{\rm RDE}$
[dotted line] is always $-1$, whereas $\tilde{\omega}_{\rm RDE}$
[solid curve] changes from 1 to $-1$ as $x$ increases from the past
($x<0$) to the future ($x>0$).}
\end{figure*}
%%%%%%%%%%%%%%%%%%%%%%%%%%%%%%%%%%%%%%%%%%%%%%%%%%%%%%%%%%%%%%%%%%%%%%%%%
Importantly, its equation of state takes the form \be
\tilde{\omega}_{\rm RDE}\equiv \frac{\tilde{p}_{\rm
RDE}}{\tilde{\rho}_{\rm
RDE}}=\frac{\frac{1-\alpha(\alpha+1)}{\alpha(\alpha+1)}\rho_{\theta
0}e^{-6x}+(1-\frac{2}{\alpha})\tilde{c}_1e^{-(4-\frac{2}{\alpha})x}}
{\frac{1-\alpha(\alpha+1)}{\alpha(\alpha+1)}\rho_{\theta
0}e^{-6x}+3\tilde{c}_1 e^{-(4-\frac{2}{\alpha})x}}. \ee For $\alpha
\simeq 1/2$ and $x>0$, we have an approximate constant equation of
state \be \tilde{\omega}_{\rm RDE}\simeq
\frac{1}{3}\Big(1-\frac{2}{\alpha}\Big) \to -1 \ee which describes
the dark energy-dominated universe deriving by cosmological constant
in the future. In order to compare  $\omega_{\rm RDE}$ with
$\tilde{\omega}_{\rm RDE}$, see Fig. 1. In this case of $
(4-2/\alpha)x \to$ const, the Friedmann equation (\ref{sfeq}) takes
an approximated from \be H^2 \approx \tilde{c}_1\ee which provides
the de Sitter-like solution \be a(t)\approx
e^{\sqrt{\tilde{c}_1}~t}. \ee Also, this form could be recovered
from (\ref{scale}) for  $t\gg1$ as \be a(t) \approx e^{\sqrt{c}_1
~t}. \ee

 On the other hand, the
modified Chaplygin gas model is given by the exotic equation of
state of  $p=-A/\rho^\alpha$ with $A>0$ and $0\le \alpha \le 1$.
Here we discuss two saturation bounds only. For $\alpha=1$, it
provides the Chaplygin gas model whose energy density is given by
\be \label{cha1} \rho_{\rm
\alpha=1}=\sqrt{A+\frac{B}{a^6}}=\sqrt{A}\sqrt{1+\frac{Be^{-6x}}{A}},
\ee where $B$ is the integration constant~\cite{Kamenshchik:2001cp}.
For $Be^{-6x}/A \gg 1$, we can approximate $ \rho_{\rm \alpha=1}$
like as \be \rho_{\rm \alpha=1}\approx \frac{\sqrt{B}}{\sqrt{A}}
e^{-3x}, p_{\rm \alpha=1}\approx 0\ee which describes the dust
matter-dominated universe with $\omega_{\rm \alpha=1}=0$ in the
early stage of the universe. For $Be^{-6x}/A \ll 1$, we have the
approximated from \be \rho_{\rm \alpha=1}\approx \sqrt{A}, p_{\rm
\alpha=1}\approx -\sqrt{A},\ee which describes the dark
energy-dominated universe $\omega_{\rm \alpha=1}=-1$ in the future.

 For $\alpha=0$ modified
Chaplygin gas model~\cite{Fabris:2003ks}, its energy density takes
\be \label{cha0}\rho_{\alpha=0}=A+A_0 e^{-3x} \ee which shows  a
dust matter-dominated phase for $x<0$, while it indicates de Sitter
phase deriving by cosmological constant for $x>0$.
%%%%%%%%%%%%%%%%%%%%%%%%%%%%%%%%%%%%%%%%%%%%%%%%%%%%%%%%%%%%%%%%%%%%%%%%%
\begin{figure*}[t!]
   \centering
   \includegraphics{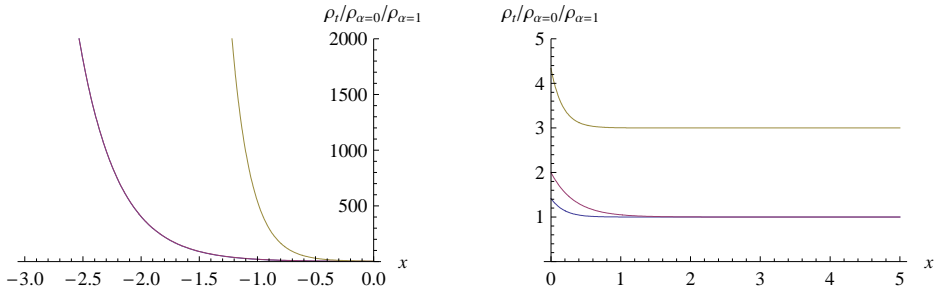}
\caption{Three energy densities as functions of $x$ for
$A=B=A_0=\rho_{\tilde{\theta}0}=\tilde{c}_1=1$ and $\alpha=1/2$. $x
= \ln[a/a_0]$ with $a_0=1$  ranges from $-\infty$ to $\infty$ which
includes the present $ x = 0$ at $a = a_0$. On the $\rho$-axis of
left-panel, from top to bottom, the curves represent
$\rho_t,\rho_{\rm \alpha=0},$ and $\rho_{\rm \alpha=1}$,
respectively. Even though all curves converge on constants for $x>0$
[right-panel] which represents de Sitter phase, their past energy
densities [left-panel] show  different behaviors for $x<0$. In this
choice of parameters, we note that $\rho_{\rm \alpha=0} \simeq
\rho_{\rm \alpha=1}$. }
\end{figure*}
%%%%%%%%%%%%%%%%%%%%%%%%%%%%%%%%%%%%%%%%%%%%%%%%%%%%%%%%%%%%%%%%%%%%%%%%%

Let us compare the total energy density (\ref{rhot}) with
(\ref{cha1}) and (\ref{cha0}).  From Fig. 2, we observe that  their
past evolutions appear differently for $x<0$, even though they
converge on  constants for $x>0$.  The (modified) Chaplygin gas
model describes the whole evolution starting from the dust
matter-dominated universe with $\omega_{\rm \alpha=0,1}=0$
 to the  dark energy-dominated universe with $\omega_{\rm \alpha=0,1}=-1$, while the Ricci dark energy
in Chern-Simons modified gravity describes the whole evolution
starting from the stiff matter-dominated universe with
$\tilde{\omega}_{\rm RDE}=1$ to the  dark energy-dominated universe
with $\tilde{\omega}_{\rm RDE}=-1$ as is depicted in Fig. 1.

Consequently, the claim of Ref.~\cite{Silva:2013yaa}  that there is
a correspondence between the Ricci dark energy in Chern-Simons
modified gravity and the modified Chaplygin gas model might be led
to misleading.  Aside from the fact that Ricci dark energy in
Chern-Simons modified gravity is nothing but Ricci dark energy with
a minimally coupled scalar when choosing the FRW metric, the
similarity between two is  limited to the de Sitter phase derived by
the cosmological constant in the future ($x>0$). This similarity can
be understood partly by reconstructing the Chaplygin gas model in
terms of the scalar~\cite{Kamenshchik:2001cp}.  The Chern-Simons
term will participate in the cosmological evolution when choosing
the anisotropic metric instead of the isotropic FRW metric
(\ref{sle})~\cite{Myung:2010qg}.

\section*{Acknowledgement}

This work was supported  by the National Research Foundation of
Korea (NRF) grant funded by the Korea government (MEST)
(No.2012-R1A1A2A10040499).

\end{document}